\documentclass[onecolumn]{emulateapj}

\usepackage[bookmarks,breaklinks]{hyperref}
   \hypersetup{
     colorlinks,
     citecolor=blue,
     linkcolor=blue,
    }

\usepackage{graphicx,times}             
\usepackage{rotating}

\shorttitle{$Insight$-HXMT study of the timing properties of Sco X-1}
\shortauthors{Jia et al.}

\received{}
\revised{}
\accepted{}

\begin{document}

\title{$Insight$-HXMT study of the timing properties of Sco X-1}

\author{
S. M. Jia$^{1,2}$, Q. C. Bu$^{1,3}$, J. L. Qu$^{1,2}$, F. J. Lu$^{1}$, S. N. Zhang$^{1,2}$,
Y. Huang$^{1,2}$, X. Ma$^{1}$, L. Tao$^{1}$, G. C. Xiao$^{1,2}$, W. Zhang$^{1,2}$
% ...
\\
(The $Insight$-HXMT Collaboration)
}

\altaffiltext{1}{Key Laboratory of Particle Astrophysics, Institute of High Energy Physics, Chinese Academy of Sciences, Beijing 100049, China}
\altaffiltext{2}{University of Chinese Academy of Sciences, Chinese Academy of Sciences, Beijing 100049, China}
\altaffiltext{3}{Institut f\"ur Astronomie und Astrophysik, Kepler Center for Astro and Particle Physics, Eberhard Karls Universit\"at, 72076 T\"ubingen, Germany}
\altaffiltext{4}{Department of Astronomy, Beijing Normal University, Beijing 100088, China}
\altaffiltext{5}{Department of Astronomy, Tsinghua University, Beijing 100084, China}
\altaffiltext{6}{Department of Physics, Tsinghua University, Beijing 100084, China}
\altaffiltext{7}{Department of Engineering Physics, Tsinghua University, Beijing 100084, China}
\altaffiltext{8}{Key Laboratory of Space Astronomy and Technology, National Astronomical Observatories, Chinese Academy of Sciences, Beijing 100012, China}
\altaffiltext{9}{College of physics Sciences \& Technology, Hebei University, Baoding 071002, Hebei Province, China}

\email{jiasm@ihep.ac.cn, qujl@ihep.ac.cn}

\begin{abstract}
We present a detailed timing study of the brightest persistent X-ray source Sco X-1 using the data collected by the Hard X-ray Modulation Telescope ($Insight$-HXMT) from July 2017 to August 2018. A complete $Z$-track hardness-intensity diagram (HID) is obtained. The normal branch oscillations (NBOs) at $\sim$ 6 Hz in the lower part of the normal branch (NB) and the flare branch oscillations (FBOs) at $\sim$ 16 Hz in the beginning part of the flaring branch (FB) are found with the data from the observations with the Low Energy X-ray Telescope (LE) and the Medium Energy X-ray Telescope (ME) of $Insight$-HXMT, while the horizontal branch oscillations (HBOs) at $\sim$ 40 Hz and the kilohertz quasi-periodic oscillations (kHz QPOs) at $\sim$ 800 Hz are found simultaneously up to 60 keV for the first time on the horizontal branch (HB) by the High Energy X-ray Telescope (HE) and ME.
We find that for all types of the observed QPOs, the centroid frequencies are independent of energy, while the root mean square (rms) increases with energy; the centroid frequencies of both the HBOs and kHz QPOs increase along the $Z$-track from the top to the bottom of the HB; and the NBOs show soft phase lags increasing with energy. A continuous QPO transition from the FB to NB in $\sim$ 200 s are also detected.
Our results indicate that the non-thermal emission is the origin of all types of QPOs, the inner most region of the accretion disk is non-thermal in nature, and the corona is nonhomogeneous geometrically.

\end{abstract}


\section{Introduction} \label{sec:intro}

X-ray binary (XRB) is a system with a central compact object (black hole or neutron star) accreting matter from its companion star. Based on the masses of the companion stars, there are two types of XRBs, the low-mass X-ray binaries (LMXBs) and high-mass binaries (HMXBs) (\citealt{Hasinger+Klis+1989}).

Bright neutron star (NS) LMXBs are further classified as $atoll$ sources and $Z$ sources, who trace distinguished tracks on the corresponding hardness-intensity diagrams (HIDs; \citealt{Hasinger+Klis+1989}) and the color-color diagrams (CCDs). $Atoll$ sources show three main states: the island, the lower banana and the upper banana states, while the $Z$ sources show a typical $Z$-shaped track with three branches: the horizontal branch (HB), the normal branch (NB) and the flaring branch (FB). Accordingly, three distinguished types of quasi-periodic oscillations (QPOs) are defined on each branch of a $Z$ source, i.e, the horizontal branch oscillation (HBO), the normal branch oscillation (NBO) and the flaring branch oscillation (FBO) (\citealt{Hasinger+Klis+1989}).
Additionally, the $Z$ sources are divided into two subgroups (\citealt{Kuulkers+etal+1994}), Cyg-like and Sco-like, and a Sco-like source has a more vertical HB and a stronger FB. In this case, the $Z$ track looks like a $v$-shape.

As the representative of the Sco-like Z-sources, Sco X-1, the brightest known persistent X-ray source (\citealt{Giacconi+etal+1962}), is a low magnetic neutron star binary ($\sim$ 1.4 $M_{\odot}$) with an $M$-class companion star ($\sim$ 0.4 $M_{\odot}$) (\citealt{Steeghs+Casares+2002}). It has QPOs observed along all the branches of its HID (\citealt{Hasinger+Klis+1989}, \citealt{Hertz+etal+1992}, \citealt{Dieters+Klis+2000}). The temporal properties of Sco X-1 have been extensively studied and well documented by many authors. Results from the data of European Space Agency's X-ray Observatory ($EXOSAT$) indicate that the QPO property is correlated with its spectral states. The QPO frequency stays in 6 -- 8 Hz on the NB, increases from $\sim$ 6 Hz to $\sim$ 16 Hz from the NB to FB, and the QPO signal disappears eventually on the upper part of the FB. Based on the $Rossi$ X-ray Timing Explorer ($RXTE$) data, \cite{Klis+etal+1996} and \cite{Yu+2007} discovered $\sim$ 45 Hz QPOs in the middle of the NB, and detected a pair of kilohertz quasi-periodic oscillations (kHz QPOs) with an upper frequency of $\sim$ 1100 Hz and a lower frequency of $\sim$ 800 Hz. \cite{Casella+etal+2006} found a smooth decrease of QPO frequency from $\sim$ 16 Hz to $\sim$ 6 Hz along with the state transition from the FB to NB.

Spectral-timing correlated studies of Sco X-1 have provided insights of how QPO properties evolve with source states, i.e., the energy dependence of the QPO properties (such as root mean square (rms), coherence and time lag) on the position of the corresponding CCDs or HIDs (\citealt{Salvo+etal+2003}, \citealt{Altamirano+etal+2008}, \citealt{Troyer+etal+2018}). Different types of QPOs may have different origins, thereby presenting various properties. Sco X-1 has been extensively studied for years, yet whose properties depending on the energy for different types of QPOs remain partly uncovered.
Due to the limited effective area of the $RXTE$ in the hard X-ray band and the relatively soft spectrum of Sco X-1, it was difficult to extend the QPO studies of this source to energies higher than 30 keV.
Thanks to the launch of $Insight$-HXMT, the Hard X-ray Modulation Telescope (\citealt{Zhang+etal+2018}, \citealt{Li+etal+2018}, \citealt{Jia+etal+2018}, \citealt{Zhang+etal+2019}), we are allowed to perform a detailed timing analysis in a broad energy band.

$Insight$-HXMT is China's first X-ray astronomical satellite, which was launched on June 15th, 2017.
It consists of three main instruments: the High Energy X-ray Telescope (HE, 20-250 keV, 5000 cm$^{2}$, 1.1$^\circ$$\times$5.7$^\circ$, $\sim$ 2 us), the Medium Energy X-ray Telescope (ME, 5-30 keV, 952 cm$^{2}$, 1$^\circ$$\times$4$^\circ$, $\sim$ 276 us), and the Low Energy X-ray Telescope (LE, 1-15keV, 384 cm$^{2}$, 1.6$^\circ$$\times$6$^\circ$, $\sim$ 1 ms).
The broad energy band, large detection area at hard X-rays and non-pile-up capability in observing bright sources make $Insight$-HXMT an ideal satellite for temporal and spectral studies of bright XRBs.

In this paper, we perform a detailed timing analysis of Sco X-1 based on the $Insight$-HXMT data from its first year observation. We search QPOs along its $Z$-track and investigate the QPO properties in different energy bands. In Section 2, we present the data selection and the data reduction processes. In Section 3, we provide the results, in terms of QPOs observed on each branch of the HID. In Section 4, we discuss the significance of our results. We give our conclusions in Section 5.

\section{Observation and Data Reduction} \label{sec:obs}

$Insight$-HXMT was targeted at Sco X-1 for nine times between July 2017 and August 2018, as listed in Table~\ref{Tab:table1}. The first observation (P0101328001) was taken during the in-orbit test, only with HE and ME turned on.
For the rest eight observations, all three instruments worked well. Each observation lasted for several satellite orbits, and was interrupted by Earth occultation and South Atlantic Anomaly (SAA).

The data reduction is performed under the $Insight$-HXMT Data Analysis Software (HDAS) V2.0 \footnote{http://www.hxmt.org/index.php/dataan/fxrj/328-data-anslysis-software-v2-update}. First, the $hepical$, $mepical$ and $lepical$ are applied to make the energy conversion for HE, ME and LE data respectively. Secondly, the $hegtigen$, $megtigen$, and $legtigen$ are applied to generate the good time intervals (GTIs), with the filtering criteria of ``$ELV>$10, $DYE\_ELV>$30, $COR>$8, $SAA\_FLAG$=0, $T\_SAA>$300, $TN\_SAA>$300 and $ANG\_DIST<$0.04", where $ELV$ is the elevation angle, $DYE$\_$ELV$ the elevation angle for the bright earth, $COR$ the geomagnetic cutoff rigidity, and $ANG$\_$DIST$ the pointing offset angle to the source. The selected GTIs are listed in Table~\ref{Tab:table1}, and the total effective observation time is about 118.8 ks. Subsequently, the $megrade$ and $lerecon$ are performed to reconstruct the events for ME and LE. At last, $hescreen$, $mescreen$ and $lescreen$ are used to extract the single event from the small field of views (FOVs) (\citealt{Zhang+etal+2019}). We chose energy band 20 -- 60 keV for HE, 8 -- 30 keV for ME and 1 -- 10 keV for LE respectively. The background subtraction of HE, ME and LE are performed with the $hebkgmap$, $mebkgmap$ and $lebkgmap$ modules provided by the $Insight$-HXMT background group.

\begin{sidewaystable*}
\centering
\caption{List of $Insight$-HXMT observations of Sco X-1 from July 2017 to August 2018, indicating the observation time, duration, functional instruments. The QPOs observed in each observation are also listed in the table. The $L_{\rm{c}}$ represents QPO centroid frequency, $L_{\rm{w}}$ represents the FWHM of QPO and rms stands for the QPO fractional rms.} \label{Tab:table1}
 \begin{tabular}{clclcccccccccc}
  \hline\noalign{\smallskip}
Obs-ID &  Dates UT      & Duration / GTI  & Instrument      &  \multicolumn{4}{c}{NBO} & \multicolumn{4}{c}{FBO} \\
  & & (orbits / ks) &  & $L_{\rm{c}}$ (Hz) & $L_{\rm{w}}$ (Hz) & rms ($\%$) & Duration (ks) & $L_{\rm{c}}$ (Hz) & $L_{\rm{w}}$ (Hz) & rms ($\%$) & Duration (ks) \\
  \hline\noalign{\smallskip}
P0101328001  & 2017 Jul. 3     & 16 / $\sim$24.0   & HE        & -- & -- & -- & -- & -- & -- & -- & --   \\
             &                 &                   & ME        & $6.5\pm{0.1}$  & $3.0\pm{0.3}$ & $3.7\pm{0.2}$ & 4.8 &
                                                                 $16.2\pm{0.2}$ & $6.6\pm{0.7}$ & $5.5\pm{0.3}$ & 6.7  \\
P0101328002  & 2018 Feb. 6-7   & 24 / $\sim$42.2   & HE        & -- & -- & -- & -- & -- & -- & -- & --   \\
             &                 &                   & ME        & $6.4\pm{0.1}$  & $3.1\pm{0.2}$ & $3.4\pm{0.1}$ & 7.7 &
                                                                 $14.3\pm{0.4}$ & $6.5\pm{1.5}$ & $4.9\pm{0.4}$ & 1.6    \\
             &                 &                   & LE        & $6.5\pm{0.2}$  & $2.7\pm{0.5}$ & $1.5\pm{0.2}$ & 7.7 &
                                                                 $13.0\pm{0.5}$ & $10.3\pm{1.7}$& $2.8\pm{0.2}$ & 1.6  \\
P0101328003  & 2018 Mar. 3     & 6  / $\sim$8.5    & HE        & -- & -- & -- & -- & -- & -- & -- & --   \\
             &                 &              & ME   & -- & -- & -- & -- & $14.0\pm{0.4}$ & $10.4\pm{1.4}$ & $6.2\pm{0.2}$ & 1.2  \\
             &                 &              & LE   & -- & -- & -- & -- & $14.6\pm{0.4}$ & $9.6\pm{2.1}$  & $2.7\pm{0.2}$ & 1.2  \\
P0101328004  & 2018 Apr. 3     & 3  / $\sim$5.3    & HE/ME/LE  & -- & -- & -- & -- & -- & -- & -- & --   \\
P0101328005  & 2018 Apr. 23    & 3  / $\sim$4.5    & HE        & -- & -- & -- & -- & -- & -- & -- & --   \\
             &                 &                   & ME        & $5.9\pm{0.1}$  & $2.8\pm{0.4}$ & $3.7\pm{0.1}$ & 2.0 &
                                                                 $15.5\pm{0.4}$ & $6.3\pm{1.3}$ & $5.2\pm{0.4}$ & 1.1  \\
             &                 &                   & LE        & $5.8\pm{0.1}$  & $2.9\pm{0.3}$ & $2.1\pm{0.1}$ & 2.0 &
                                                                 $16.1\pm{0.8}$ & $9.2\pm{4.0}$ & $2.0\pm{0.1}$ & 1.1  \\
P0101328006  & 2018 May  27    & 3  / $\sim$4.3    & HE/ME/LE  & -- & -- & -- & -- & -- & -- & -- & --   \\
P0101328008  & 2018 Jun. 24-25 & 15 / $\sim$13.2   & HE/ME/LE  & -- & -- & -- & -- & -- & -- & -- & --   \\
P0101328009  & 2018 Jul. 26-27 & 6  / $\sim$8.3    & HE        & -- & -- & -- & -- & -- & -- & -- & --   \\
             &                 &                   & ME        & $6.7\pm{0.2}$  & $3.8\pm{0.5}$ & $3.6\pm{0.2}$ & 4.4 &
                                                                 $18.0\pm{0.5}$ & $9.5\pm{1.9}$ & $4.5\pm{0.3}$ & 1.8  \\
             &                 &                   & LE        & $6.3\pm{0.2}$  & $2.9\pm{0.5}$ & $1.7\pm{0.1}$ & 4.4 &
                                                                 $17.1\pm{0.5}$ & $8.7\pm{1.7}$ & $2.1\pm{0.2}$ & 1.8  \\
  \hline\noalign{\smallskip}
  &  &  &  &  \multicolumn{4}{c}{HBO} & \multicolumn{4}{c}{kHz QPO} \\
  &  &  &  & $L_c$ (Hz) & $L_w$ (Hz) & rms ($\%$) & Time (ks)  & $L_c$ (Hz) &  $L_w$ (Hz) & rms ($\%$) &  Time (ks) \\
  \hline\noalign{\smallskip}
P0101328010  & 2018 Aug. 16    & 6  / $\sim$8.5    & HE        & $39.5\pm{0.7}$   & $6.5\pm{3.0}$   & $11.2\pm{1.3}$ & 5.4 &
                                                                 $820.4\pm{12.3}$ & $60.0\pm{21.6}$ & $12.4\pm{1.4}$ & 5.4  \\
             &                 &                   & ME        & $43.0\pm{3.1}$   & $35.4\pm{6.3}$  & $3.8\pm{0.3}$  & 5.4&
                                                                 $807.5\pm{11.0}$ & $147.2\pm{50.0}$& $5.2\pm{0.6}$  & 5.4  \\
             &                 &                   & LE        & -- & -- & -- & -- & -- & -- & -- & --   \\
  \noalign{\smallskip}\hline
\end{tabular}
\end{sidewaystable*}

\section{Results} \label{sec:data}

\subsection{Hardness Intensity Diagram} \label{subsec:pds}

\begin{figure*}
   \centering
   \includegraphics[width=0.7\textwidth, angle=0]{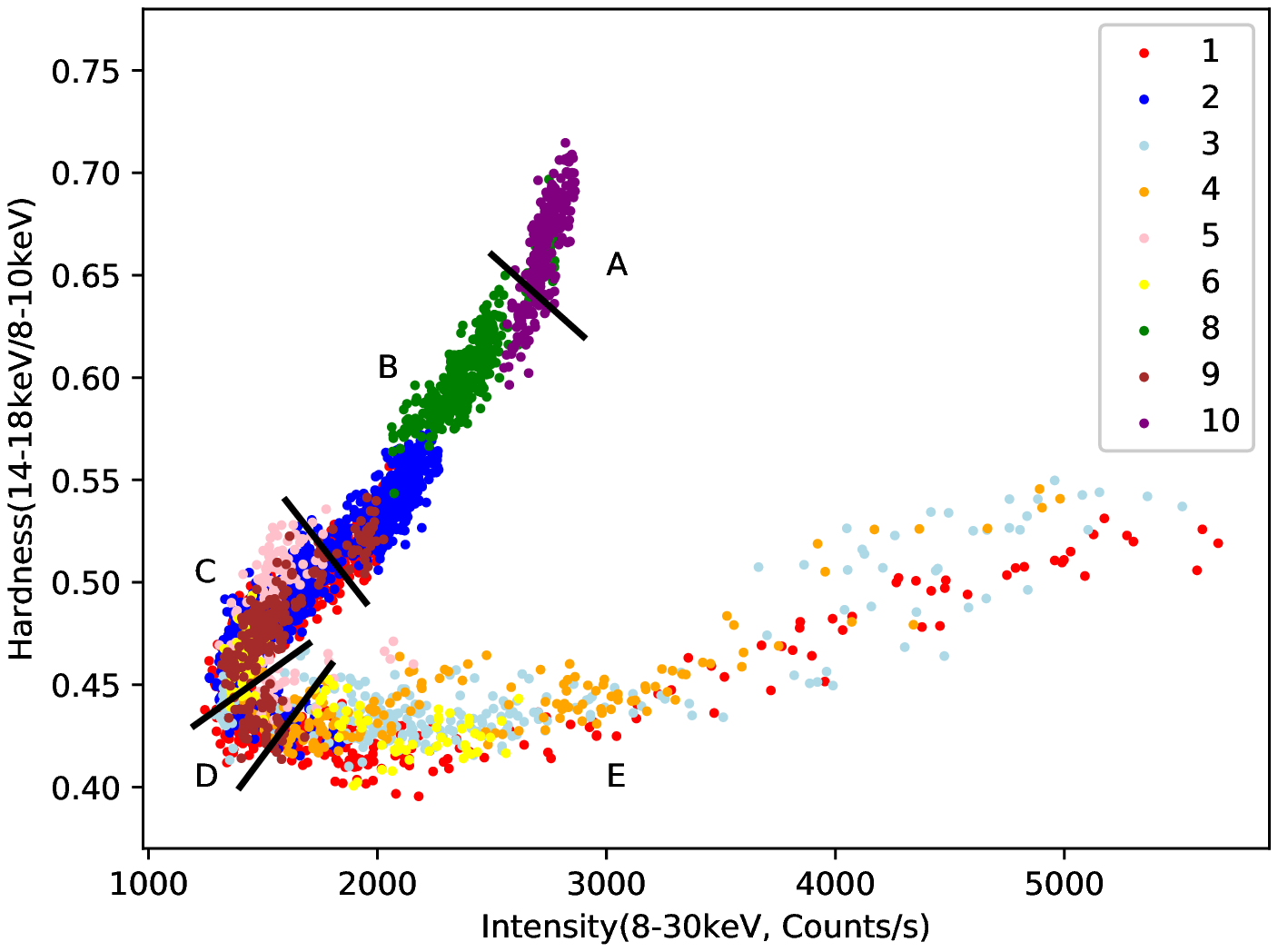}
   \caption{Hardness Intensity Diagram produced from all the $Insight$-HXMT/ME observations of Sco X-1 listed in Table\ref{Tab:table1}. Each data point corresponds to 32 s of data, and the different colors mark the different observations as denoted. Based on the distribution of QPOs, HID is divided into five regions as A, B, C, D and E. NBOs are observed in region C, FBOs are observed in region D, and HBOs and kHz QPOs are observed in region A.}
   \label{Fig:HID}
   \end{figure*}

The HID is produced from ME data. For each 32 s data segment, we define the hardness as ratio of the net count rates from two different energy bands, $\sim$ 14 -- 18 keV and $\sim$ 8 -- 10 keV. The intensity is defined as the net count rate in $\sim$ 8 -- 18 keV. The HID is plotted in Fig.~\ref{Fig:HID}, and the nine different colors distinguish each observation as illustrated in the figure. The HID of Sco X-1 shows a complete $Z$ track, which is similar to that of \cite{Titarchuk+etal+2014} made from $RXTE$ data, exhibiting a $v$-like track with a short and vertical HB. The upper left part of the $v$ track corresponds to the HB and NB, and the lower right part corresponds to the FB.
The results in \cite{Titarchuk+etal+2014} suggested that the shapes of the $Z$-tracks remained more of less the same in several years, except shifts between different tracks. In our case, a complete $v$-like shaped HID is formed, and all the observations are located on the $v$-track precisely.

\subsection{Quasi-Periodic Oscillations Along the $Z$-track} \label{subsec:d4}

In order to search for low frequency QPOs (LFQPOs), we create power density spectrum (PDS) for each observation from 32 time stretches, with a time resolution of 1/512 s (corresponding to a Nyquist frequency of 256 Hz). For the inspection of the high frequency QPOs (HFQPOs), we compute the PDS with a time resolution of 1/4096 s (corresponding to a Nyquist frequency of 2048 Hz). The PDS fitting is taken with the {\sc xspec} (v12.9.1), by applying a one-to-one energy-frequency conversion with a unit response (\citealt{Casella+etal+2004}). A three-component model is applied to fit the PDS: two Lorentzians and a power law. In particular, a Lorentzian is used to fit the very low frequency noise (VLFN), a power law to fit the Poisson noise, and another narrower Lorentzian to fit the QPO component. In some cases, an extra Lorentzian component is needed to fit the kHz QPO.

Three parameters of QPOs are obtained from the PDS fitting, i.e., the centroid frequency $L_{\rm{c}}$, the full width at half maximum (FWHM) $L_{\rm{w}}$ and the normalization measuring the signal strength $L_{\rm{n}}$. The fractional rms amplitude is calculated by the integral of the normalized Lorenzian PDS in the appropriate frequency range and taking the mean count rate of the source into account (\citealt{Lewin+etal+1988}). Considering the contribution of background, the corrected fractional rms is calculated as rms=$\sqrt{L_{\rm{n}}/(S+B)}\times(S+B)/S$ (\citealt{Bu+etal+2015}), where $S$ and $B$ stand for source and background count rates respectively. The errors of the rms are estimated with a standard error propagation (\citealt{Bevington+Robinson+2003}). The fitted results are listed in Table\ref{Tab:table1}.

We find the $\sim$ 6 Hz NBOs in four observations (P0101328001, P0101328002, P0101328005 and P0101328009) and the $\sim$ 16 Hz FBOs in five observations (P0101328001, P0101328002, P010328003, P0101328005 and P0101328009). The $\sim$ 40 Hz HBOs and the $\sim$ 800 Hz kHz QPOs are only observed in P0101328010. We note that the NBOs and FBOs are only detected by LE and ME, while the HBOs and kHz QPOs are only detected by ME and HE.
The non-detection of the kHz QPOs by LE is partly due to its limited time resolution, which is about 1 ms. But given that the rms amplitude of the kHz QPO in the ME detection is much smaller than that in the HE detection, the kHz QPO might be also undetectable in the LE energy band even if it had a high enough time resolution.

By comparing the positions of the QPOs on the HID, we find that the NBOs are observed on the lower part of the NB, the FBOs on the beginning part of the FB, and the HBOs and kHz QPOs on the HB of the $Z$-track. We thereby divide the $Z$-track into five regions as A (HB, HBOs and kHz QPOs), B (upper NB), C (lower NB, NBOs), D (lower FB, FBOs) and E (upper FB), as marked in Fig.~\ref{Fig:HID}, and then investigate the QPO properties in each region respectively.

\subsubsection{NBOs and FBOs} \label{subsec:d3}

\begin{figure*}[h]
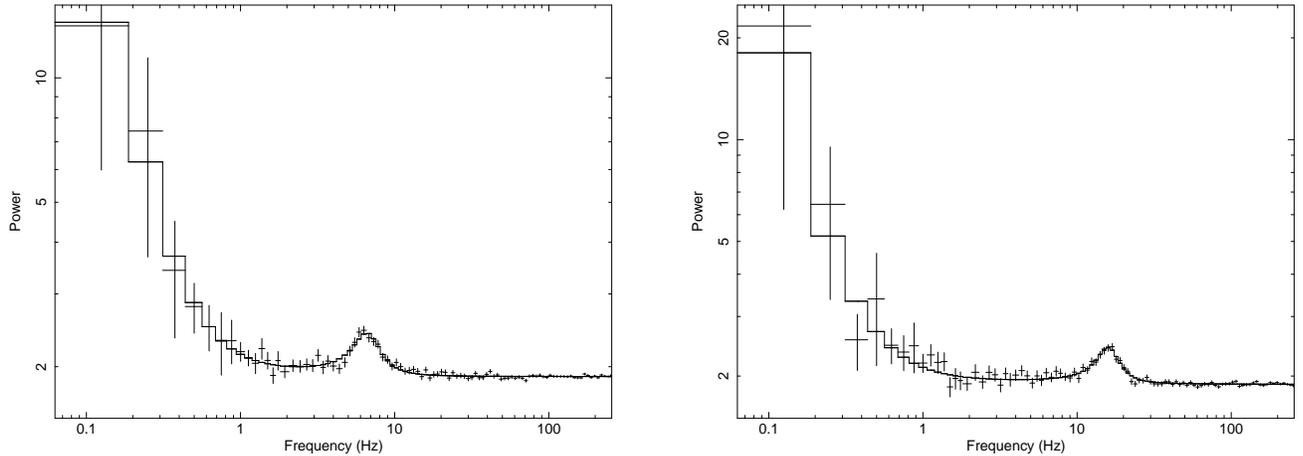

  \begin{minipage}{0.5\linewidth}
  \centering
   \includegraphics[width=60mm,angle=270]{ME_NBO.eps}
  \end{minipage}%
  \begin{minipage}{0.5\linewidth}
  \centering
   \includegraphics[width=60mm,angle=270]{ME_FBO.eps}
  \end{minipage}%
  \caption{{The average PDS of Sco X-1 computed from ME of $Insight$-HXMT, fitted by a power law and two Lorentzians components. Left panel is the PDS of the $\sim$ 6 Hz NBO from region C, and the right panel is the PDS of the $\sim$ 16 Hz FBO from region D. The offset of the continuum from a level of 2.0 is due to the dead time effects of ME.}}
  \label{Fig:QPO2}
\end{figure*}

As illustrated in Fig.~\ref{Fig:HID}, the NBOs are concentrated in region C, while the FBOs are concentrated in region D. In order to investigate the properties of NBOs and FBOs, we combine the NBO PDS in region C and the FBO PDS in region D, respectively. The PDS are fitted by the three-component model as described in the beginning part of Section 3.2. The fitted results of NBOs and FBOs are listed in Table~\ref{Tab:table2}, and the averaged ME PDS of regions C and D are shown in Fig.~\ref{Fig:QPO2}.

Table~\ref{Tab:table2} shows that the centroid frequencies of the NBOs stay constantly at $\sim$ 6.6 Hz in the LE and ME energy bands, while the centroid frequencies of FBOs increase from $13.8\pm{0.3}$ Hz (LE) to $15.8\pm{0.2}$ Hz (ME). However, when we select the same good time intervals (GTIs) for LE and ME, the frequency difference of FBOs between two instruments disappears. These results suggest that the frequencies of both the NBOs and FBOs are energy-independent but vary with time. In addition, the NBO and FBO rms of ME are larger than those of LE, suggesting that the rms of the NBOs and FBOs increase with energy.

\begin{table*}
\begin{center}
\caption[]{The fitting parameters of NBOs from region C and FBOs from region D observed by ME and LE. $L_{\rm{c}}$ represents the QPO centroid frequency, $L_{\rm{w}}$ represents the FWHM and rms is the QPO fractional rms.}\label{Tab:table2}
 \begin{tabular}{ccccccc}
  \hline\noalign{\smallskip}
              & \multicolumn{3}{c}{ME}               & \multicolumn{3}{c}{LE}                \\
              & $L_{\rm{c}}$ (Hz) & $L_{\rm{w}}$ (Hz) & rms($\%$)  & $L_{\rm{c}}$ (Hz) &  $L_{\rm{w}}$ (Hz) & rms($\%$)  \\
  \hline\noalign{\smallskip}
Region C (NBO) & $6.6\pm{0.1}$  & $3.3\pm{0.3}$ & $4.1\pm{0.2}$ & $6.4\pm{0.2}$  & $2.7\pm{0.7}$  & $1.2\pm{0.1}$  \\
Region D (FBO) & $15.8\pm{0.2}$ & $6.6\pm{0.6}$ & $6.2\pm{0.2}$ & $13.8\pm{0.3}$ & $10.7\pm{1.3}$ & $2.6\pm{0.2}$   \\
  \noalign{\smallskip}\hline
\end{tabular}
\end{center}
\end{table*}

In order to study the energy dependence of NBOs, we further selected four energy bands: 1.0 -- 2.8 keV, 2.8 -- 4.6 keV, 4.6 -- 6.4 keV and 6.4 -- 8.2 keV from LE, and three energy bands: 8.2 -- 10 keV, 10 -- 14 keV and 14 -- 30 keV from ME. We compute the PDS in these energy bands and fit them as described in the beginning part of Section 3.2.

For all the NBO observations, only P010132800202, P010132800204 and P010132800206 are statistically good enough in all seven sub-channels to identify NBOs. The quality factor ($Q$) is defined as $Q=L_{\rm{c}}/L_{\rm{w}}$ (\citealt{Barret+etal+2005}). The phase lag of the NBOs is calculated by averaging the phase lags over the frequency range from $L_{\rm{c}}-L_{\rm{w}}/2$ to $L_{\rm{c}}+L_{\rm{w}}/2$ (\citealt{Qu+etal+2010}, \citealt{Huang+etal+2018}). The reference energy band for phase lag is 1.0 -- 2.8 keV.

\begin{figure*}[h]
  \begin{minipage}[t]{0.5\linewidth}
  \centering
   \includegraphics[width=90mm,angle=0]{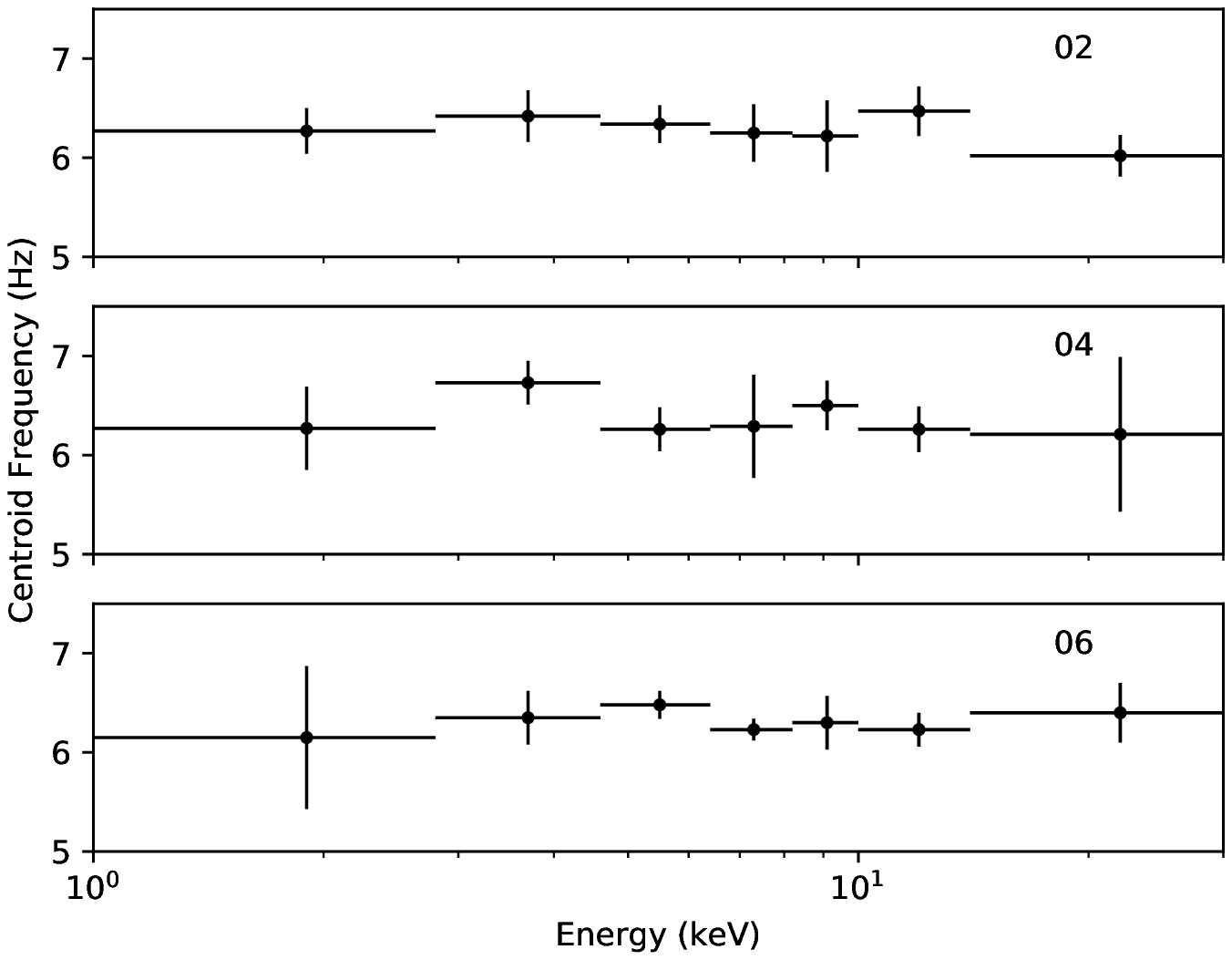}
  \end{minipage}%
  \begin{minipage}[t]{0.5\linewidth}
  \centering
   \includegraphics[width=90mm,angle=0]{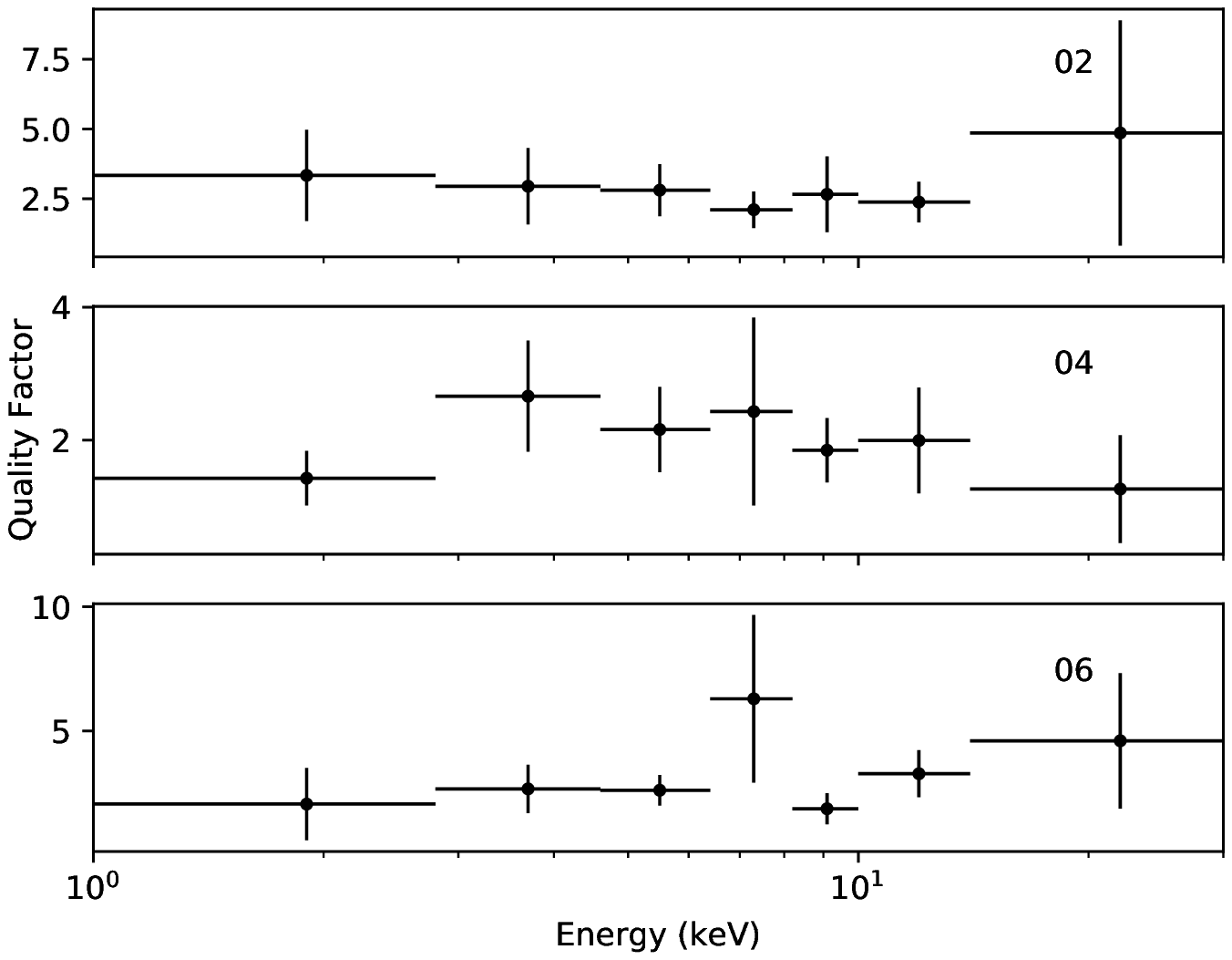}
  \end{minipage}%
  \\
  \begin{minipage}[t]{0.5\linewidth}
  \centering
   \includegraphics[width=90mm,angle=0]{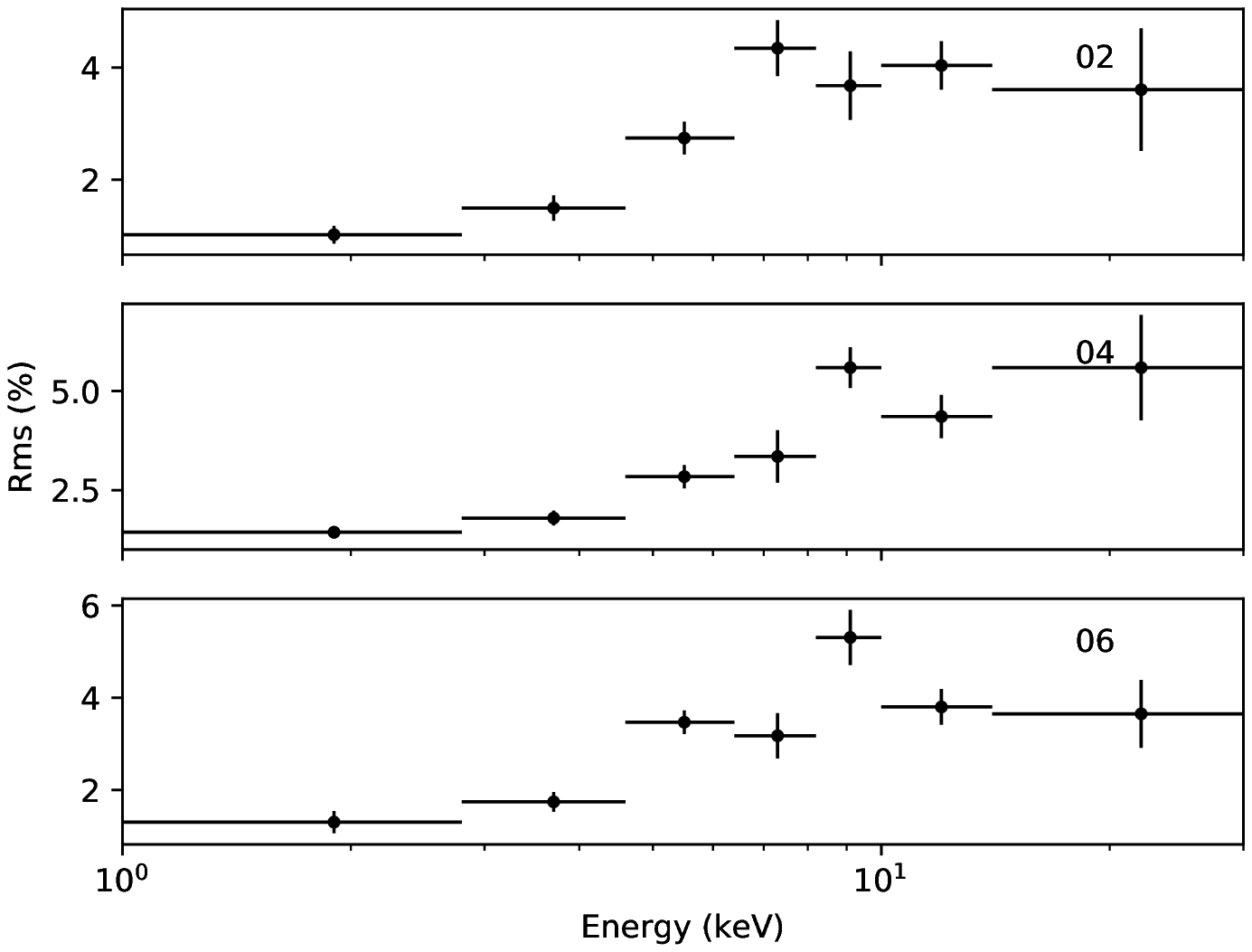}
  \end{minipage}%
  \begin{minipage}[t]{0.5\linewidth}
  \centering
   \includegraphics[width=90mm,angle=0]{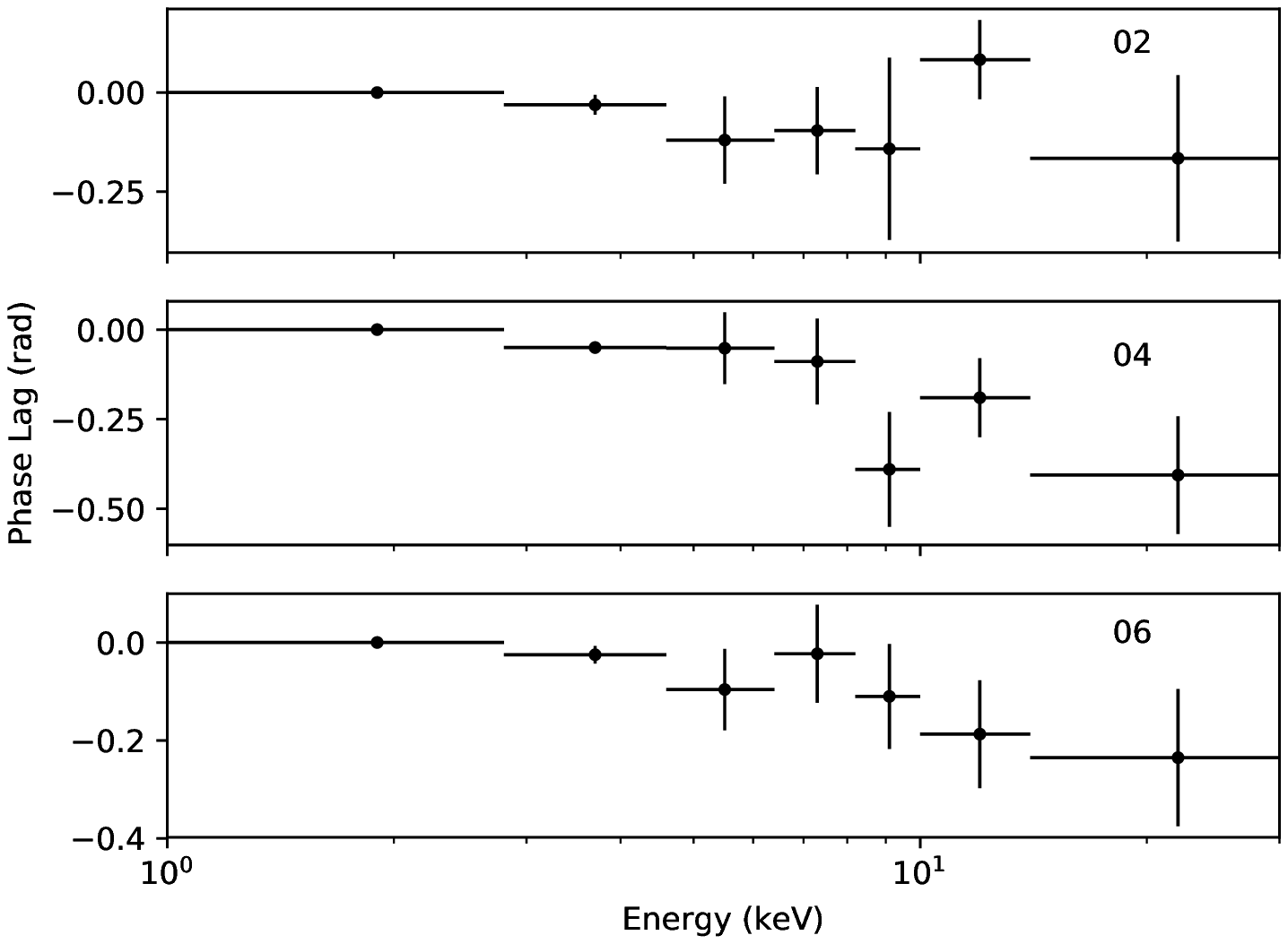}
  \end{minipage}%
  \caption{The NBO centroid frequency, quality factor, factional rms and phase lag as a function of photon energy, from left to right, top to bottom. The three sub-figures of four panels are from observations P010132800202, P010132800204 and P010132800206.}
  \label{Fig:NBO}
\end{figure*}

In Fig.~\ref{Fig:NBO}, the centroid frequencies of NBOs stay constantly with the energy increasing from 1 keV to 30 keV (the upper left panel). No significant correlation is found between the quality factor and photon energy, suggesting that the coherence of NBOs is independent of photon energy. The fractional rms increases with photon energy before reaching a plateau around $\sim$ 10 keV in the lower left panel. Despite the large errors, the NBO phase lag shows a negative correlation with photon energy.

\subsubsection{Transition from the FBO to NBO} \label{subsec:d3}

\begin{figure*}[h]
  \begin{minipage}[t]{0.5\linewidth}
  \centering
  \includegraphics[width=90mm]{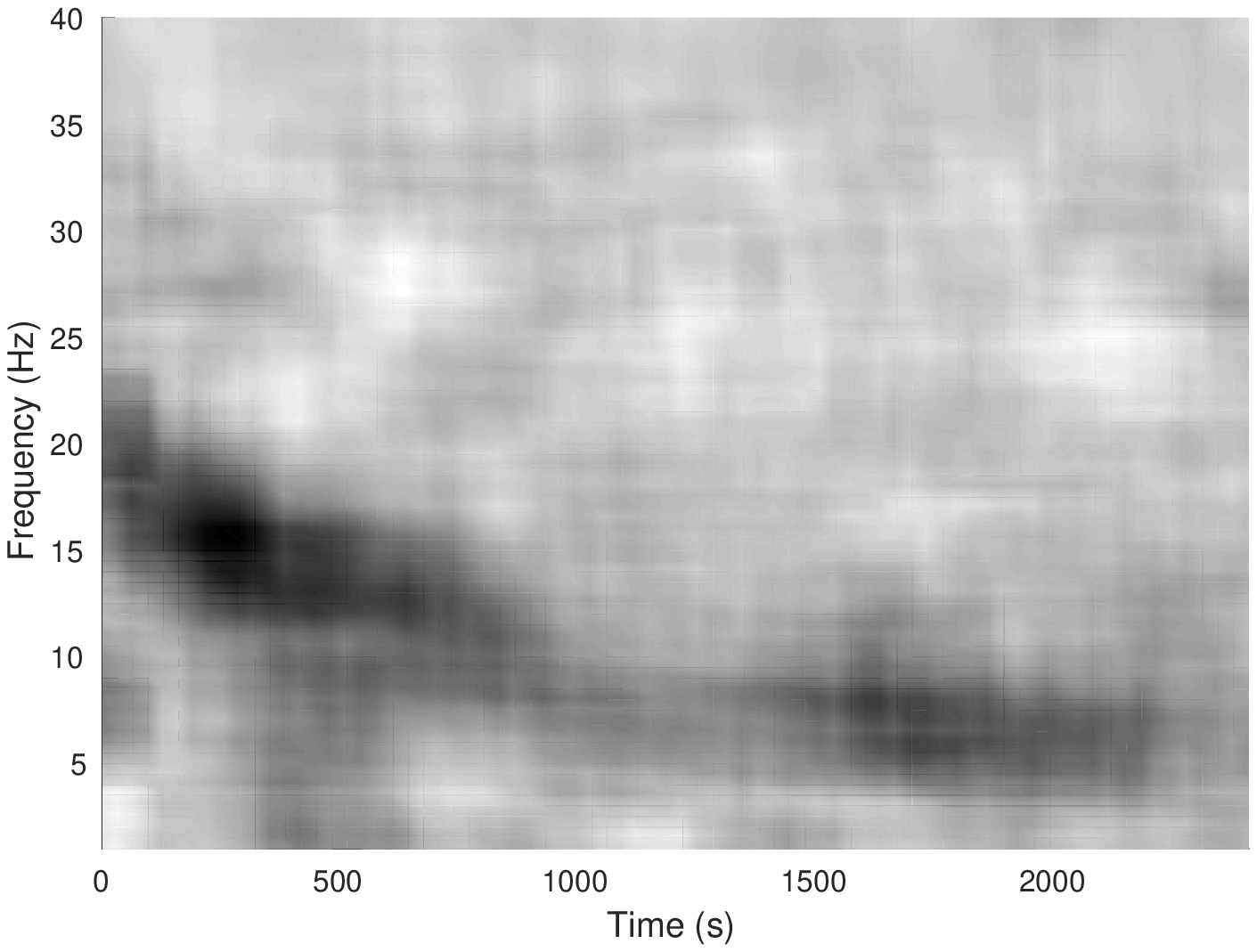}
  \end{minipage}%
  \begin{minipage}[t]{0.5\linewidth}
  \centering
  \includegraphics[width=90mm]{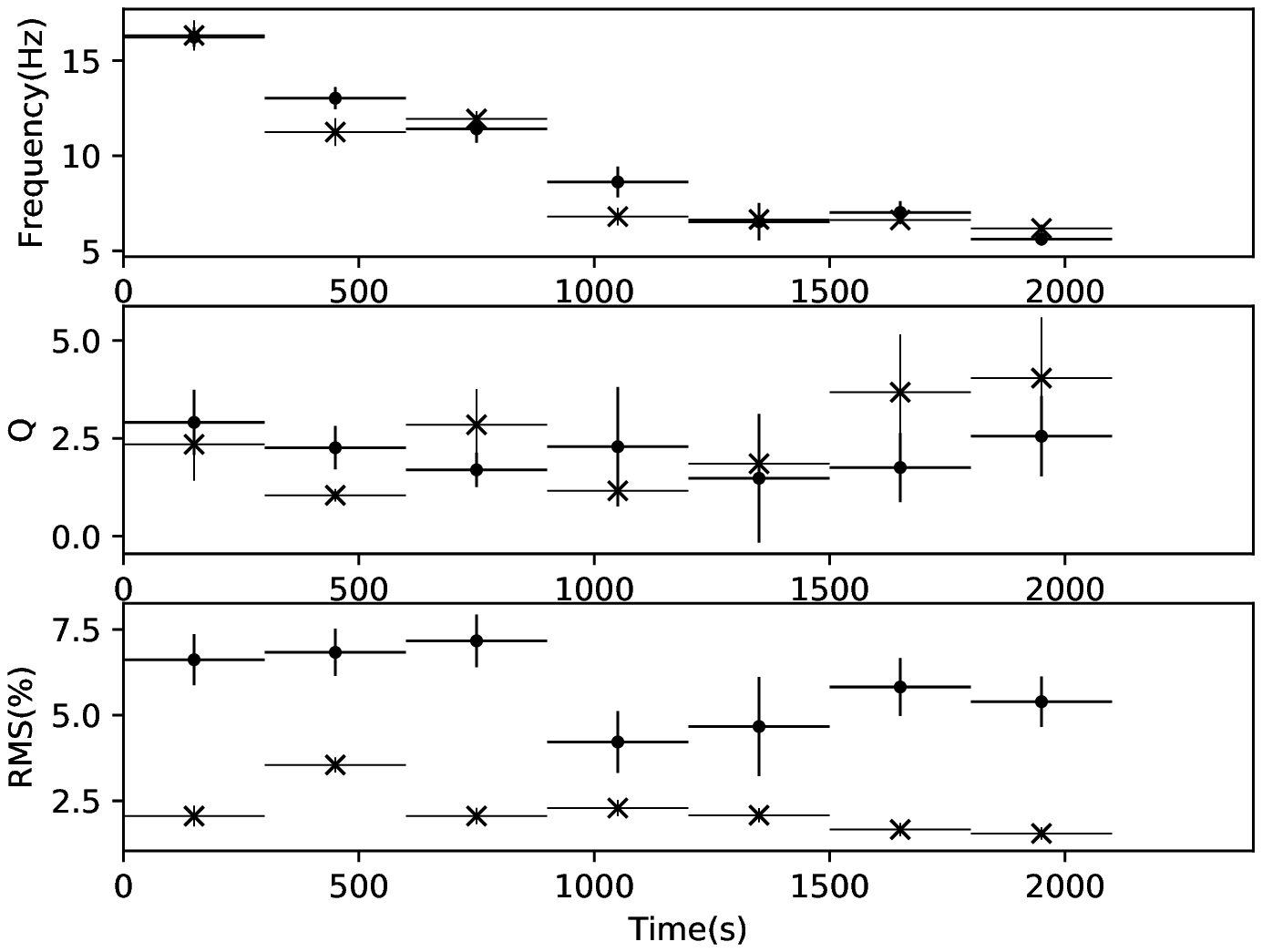}
  \end{minipage}%
  \caption{{Left: The dynamical PDS of the observation P010132800203. Right: QPO centroid frequency, quality factor and fractional rms as a function of time when the source experiences the transition from the FB to NB. The solid points and the stars represent ME and LE data, respectively.}}
  \label{Fig:FBtoNB}
\end{figure*}

A continuous QPO transition from FB to NB on a time scale of 2400 s is found in the observation of P010132800203 by ME and LE, whose dynamical PDS of ME is plotted in the left panel of Fig.~\ref{Fig:FBtoNB}. The QPO frequency starts at $\sim$ 20 Hz from the beginning and decreases to $\sim$ 15 Hz on the FB, then rapidly decreases to $\sim$ 6 Hz in $\sim$ 200 s from the FB to NB, and stabilizes at $\sim$6 Hz in the lower part of the NB for $\sim$ 1000 s and disappears in the upper part of the NB.

In order to investigate the QPO properties during the transition, seven intervals for every 300 s are further extracted from P010132800203. The results are plotted in the right panel of Fig.~\ref{Fig:FBtoNB}, in which the solid points represent ME data and the stars refer to LE data. Besides the decreasing QPO frequency, from $\sim$ 16 Hz to $\sim$ 6 Hz, the QPO quality factor is likely independent of time or QPO type within 1$\sigma$ error range. The QPO fractional rms increases with photon energy, which is consistent with the result shown in the lower left panel of Fig.~\ref{Fig:NBO}.

\subsubsection{HBOs and kHz QPOs} \label{subsec:d3}

\begin{figure*}[h]
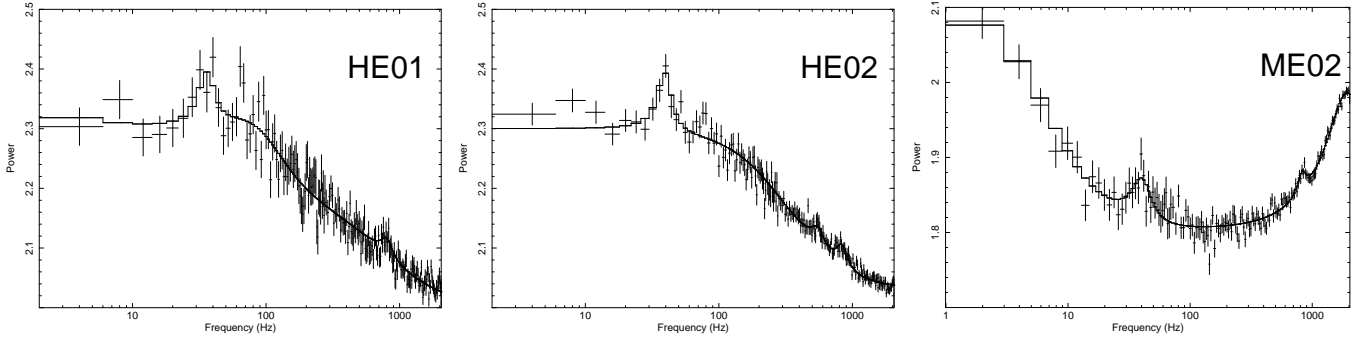

  \begin{minipage}[t]{0.33\linewidth}
  \centering
   \includegraphics[width=45mm,angle=270]{HBO_1001_HE_new.eps}
  \end{minipage}%
  \begin{minipage}[t]{0.33\linewidth}
  \centering
   \includegraphics[width=45mm,angle=270]{HBO_1002_HE_new.eps}
  \end{minipage}%
  \begin{minipage}[t]{0.33\linewidth}
  \centering
   \includegraphics[width=45mm,angle=270]{HBO_1002_ME_new.eps}
  \end{minipage}%
  \caption{{The PDS of Sco X-1 fitted by a power law and three Lorentzians components. Left: the HBO and kHz QPO detected by HE in P010132801001. Mid: the HBO and kHz QPO detected by HE in P010132801002. Right: the HBO and kHz QPO detected by ME in P010132801002. }}
  \label{Fig:HBO}
\end{figure*}

The $\sim$ 40 Hz HBOs and $\sim$ 800 Hz QPOs are found in P010132801001/HE and P010132801002/ME/HE, which correspond to the HB on the $Z$-track shown in Fig.~\ref{Fig:HID}. The corresponding PDS with best-fitting lines are plotted in Fig.~\ref{Fig:HBO}\footnote{The shape of ME PDS and the offset of the continuum from a level of 2.0 are caused by the dead time effect (\citealt{Zhang+etal+1995}), and the shape of HE PDS is due to the $spike$ correction during the data analysis.}. However, no significant HBO or kHz QPO signals are detected by LE at the same time.

The centroid frequency, FWHM, quality factor, fractional rms and significance of the HBO signals are list in Table~\ref{Tab:tableHBO}. In Table~\ref{Tab:tableHBO}, within the same observation (P010132801002), the HBO centroid frequency stays constantly from ME to HE at $\sim$ 40.0 Hz. While moving from the top of the HB (P010132801001) to the bottom of the HB (P010132801002), the HBO frequency increases slightly from $\sim$ $36.0^{+2.2}_{-3.2}$ Hz to $\sim$ $39.4^{+1.0}_{-0.9}$ Hz.

\begin{table*}
\begin{center}
\caption{The HBOs detected in the observation of P0101328010 by ME and HE.}\label{Tab:tableHBO}
 \begin{tabular}{clclclclcl}
  \hline\noalign{\smallskip}
Obs-ID      & Instrument & $L_{\rm{c}}$ (Hz)      & $L_{\rm{w}}$ (Hz)       & Q &  rms($\%$)      & S/N       \\
  \hline\noalign{\smallskip}
P010132801001   &  HE    &  $36.0^{+2.2}_{-3.2}$    &  $12.2^{+4.6}_{-3.0}$    & 3.0$\pm$1.1 & 7.7$\pm$1.1   &$\sim3.4\sigma$  \\
P010132801002   &  HE    &  $39.4^{+1.0}_{-0.9}$    &  $11.5^{+5.5}_{-5.4}$    & 3.4$\pm$1.2 & 7.9$\pm$0.7   &$\sim5.6\sigma$  \\
                &  ME    &  $40.0^{+3.2}_{-1.3}$    &  $18.8^{+7.0}_{-2.6}$    & 2.1$\pm$0.6 & 3.0$\pm$0.3   &$\sim5.4\sigma$  \\
  \noalign{\smallskip}\hline
\end{tabular}
\end{center}
\end{table*}

\begin{figure*}[h]
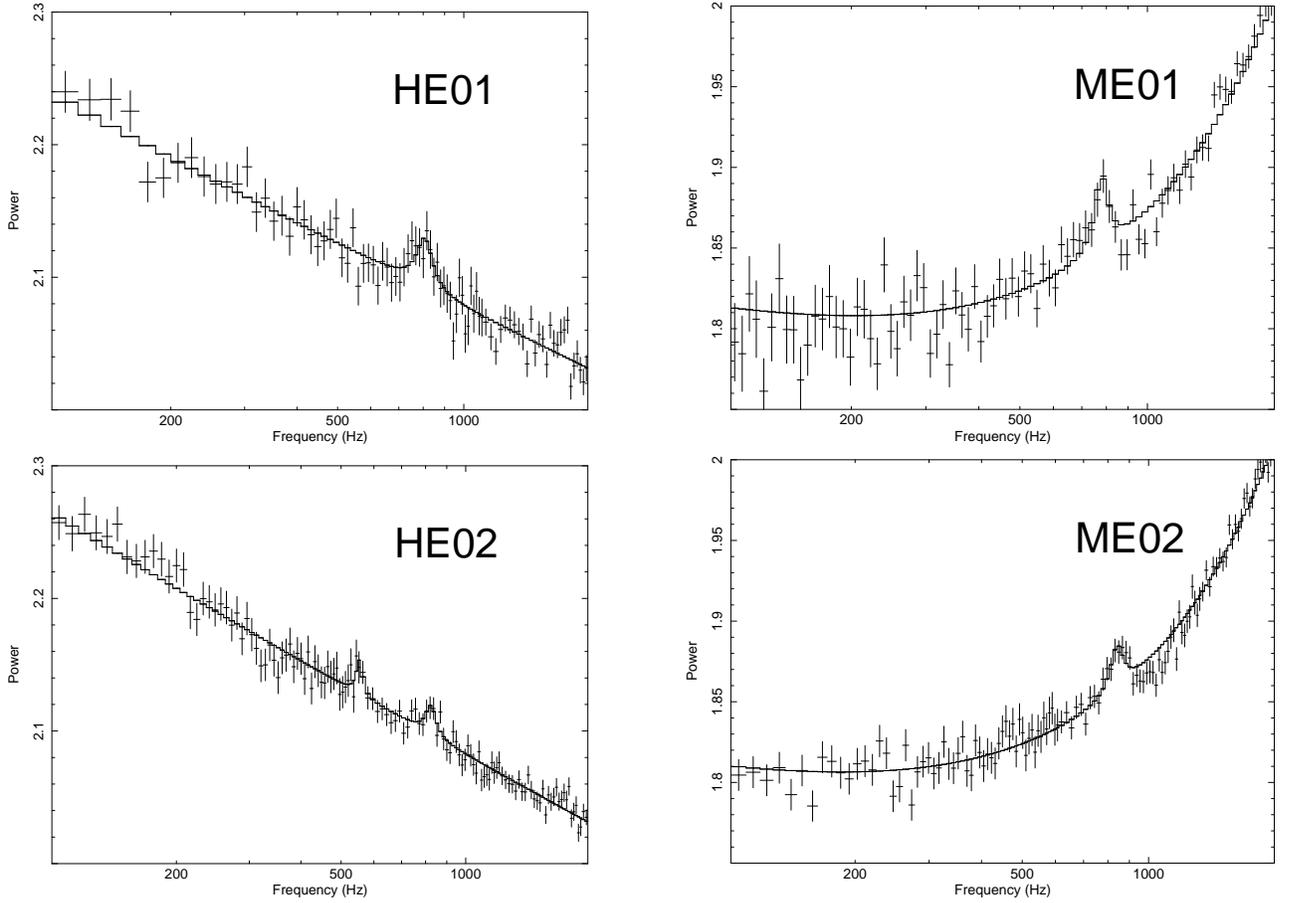

  \begin{minipage}[t]{0.5\linewidth}
  \centering
   \includegraphics[width=60mm,angle=270]{kHzQPO_1001_HE.eps}
  \end{minipage}%
  \begin{minipage}[t]{0.5\linewidth}
  \centering
   \includegraphics[width=60mm,angle=270]{kHzQPO_1001_ME.eps}
  \end{minipage}%
  \\
  \begin{minipage}[t]{0.5\linewidth}
  \centering
   \includegraphics[width=60mm,angle=270]{kHzQPO_1002_HE.eps}
  \end{minipage}%
  \begin{minipage}[t]{0.5\linewidth}
  \centering
   \includegraphics[width=60mm,angle=270]{kHzQPO_1002_ME.eps}
  \end{minipage}%
  \caption{{The PDS of Sco X-1 fitted by a power law and two Lorentzians components. Upper: the kHz QPOs detected by HE and ME in P010132801001. Lower: the kHz QPOs detected by HE and ME in P010132801002. }}
  \label{Fig:kHzQPO}
\end{figure*}

In order to study the properties of the kHz QPOs, we rebin the PDS and fit them in the frequency range of 100-2000 Hz, as shown in Fig.~\ref{Fig:kHzQPO}.
The centroid frequency, FWHM, quality factor, fractional rms and significance of the kHz QPOs are listed in Table~\ref{Tab:tablekHzQPO}. Taking the error bars into consideration, the kHz QPO centroid frequencies show no variation with energy up to 60 keV, but increase from the top of the HB (P010132801001) to the bottom of the HB (P010132801002) along the $Z$-track. In addition, the rms of HBOs and kHz QPOs increase from ME to HE, from $\sim 5 \%$ to $\sim 14 \%$

\begin{table*}
\begin{center}
\caption[]{ The kHz QPOs detected in the observation of P0101328010 by ME and HE.}\label{Tab:tablekHzQPO}
 \begin{tabular}{clclclclcl}
  \hline\noalign{\smallskip}
Obs-ID      & Instrument & $L_{\rm{c}}$ (Hz)      & $L_{\rm{w}}$ (Hz)       & Q &  rms($\%$)      & S/N       \\
  \hline\noalign{\smallskip}
P010132801001   &  HE    &  $804.5^{+10.4}_{-18.1}$   &  $91.8^{+37.7}_{-26.1}$    & 8.8$\pm$3.0  & 13.9$\pm$0.7   &$\sim3.9\sigma$   \\
                &  ME    &  $778.4^{+16.8}_{-12.7}$   &  $68.2^{+18.7}_{-58.4}$    & 11.4$\pm$6.4 & 5.2$\pm$0.6   &$\sim4.1\sigma$  \\
P010132801002   &  HE    &  $824.4^{+9.5}_{-10.8}$    &  $60.5^{+32.4}_{-29.9}$    & 13.6$\pm$6.9 & 9.2$\pm$1.5   &$\sim3.0\sigma$  \\
                &  ME    &  $840.3^{+7.9}_{-10.2}$    &  $72.9^{+15.3}_{-63.2}$    & 11.5$\pm$6.2 & 4.2$\pm$0.5   &$\sim5.3\sigma$  \\
  \noalign{\smallskip}\hline
\end{tabular}
\end{center}
\end{table*}

\section{DISCUSSION} \label{sec:data}

In the previous section we have obtained various QPO behaviors in Sco X-1. The full coverage of the $Z$-track and the broad energy band of $Insight$-HXMT provide us an opportunity to study the evolution of QPOs with photon energy and spectral states. For all types of QPOs (NBO, FBO, HBO and kHz QPO), their centroid frequencies are all energy independent, while their rms increases with energy. For HBOs and kHz QPOs, their centroid frequencies increase along the $Z$-track from the top to the bottom of the HB. In the following we compare our results with the previous ones and discuss about their physical implications.

\subsection{Energy-dependence of QPOs}

In Section 3.2.1, we studied the energy dependence of NBOs in seven energy bands within $\sim$ 1.0 -- 30.0 keV. The NBOs found here have frequencies similar to what has been reported in \cite{Wang+etal+2012}, with a value of $\sim$ 6.2 -- 6.5 Hz (Fig.~\ref{Fig:NBO}). In their work, they proposed that NBO centroid frequency evolved with photon energy from a negative correlation to a positive one within a frequency interval of $\nu_0\pm0.3$ Hz. However, considering the error bars of our results, we suggest that the NBO centroid frequencies do not vary with photon energy at 68\% confidence ($\Delta\nu/\nu_0<$ 4.0\%).

The quality factor of NBO shows no obvious correlation with photon energy, while the NBO fractional rms increases with photon energy before reaching a plateau at $\sim$ 10 keV, which are consistent with that reported in \cite{Wang+etal+2012},
and suggest that the NBOs may originate from the non-thermal emission.

The phase lags of NBOs in different energy bands in Sco X-1 are also investigated in this work (Fig.~\ref{Fig:NBO}). The soft time lag reaches its maximum of $\sim$ 8 ms between 14 -- 30 keV and 1.0 -- 2.8 keV.
Normally, the millisecond time lag can be explained by the process of Compton up-scattering, in which the seed soft photons are up-scatted to higher energies by the high energy electrons in the corona. However, to produce higher energy photons more scatters are needed, and thus hard time lags are expected, in contradiction to our observations. On the other side, the timescale of the lag we obtained ($\sim$ 8 ms) is too short to be explained by the propagating fluctuations model, in which a $\sim$ 1/$f$ frequency-dependence of time lags ($\sim$ 1/6.5 s for NBOs) is expected (\citealt{Arevalo+Uttley+2006}, \citealt{Uttley+etal+2011}).
\cite{Nobili+etal+2000} proposed a non-homogeneous Compton cloud model to explain the soft time lags in GRS 1915+105. In their model, the seed photons are up-scattered in the inner region of the corona that has higher temperature and large scattering depths than the outer part of corona. The soft lag are thereby caused by the photons down-scattered in the cooler and outer part of the corona.
Therefore, such a non-homogeneous corona can induce both hard and soft time lags.
In this scenario, the time lag, rms, frequency correlations with photon energy detected by $Insight$-HXMT can be naturally explained.

\subsection{The FBO-to-NBO transition} \label{subsec:d3}

The QPO transition of the FBO to NBO along the $Z$-track in Sco X-1 has been studied in several works. \cite{Priedhorsky+etal+1986} first reported a transition from $\sim$ 21 Hz FBO to $\sim$ 6 Hz NBO in about 1000 s in this source with $EXOSAT$. A dramatic jump in the QPO frequency at the transition from the NB $\sim$ 7 Hz to FB $\sim$ 17 Hz was detected by \cite{Dieters+Klis+2000}, but the actual transition was not resolved. \cite{Casella+etal+2006} reported a clear continuous fast transition from $\sim$ 16 Hz FBO to $\sim$ 6 Hz NBO on a time scale of $\sim$ 100 s with $RXTE$. \cite{Titarchuk+etal+2014} also reported a transition of $\sim$ 15 Hz FBO to $\sim$ 6 Hz NBO with $RXTE$. Here we found a QPO transition from $\sim$ 15 Hz to $\sim$ 6 Hz in about 200 s (from 700 s to 900 s in Fig.~\ref{Fig:FBtoNB}) with $Insight$-HXMT. Therefore, the typical time scale of the transition from FBO to NBO is probably 100 -- 200 s.  The QPO frequency changes over the FB and its absence at the top of the FB are found to be related with the spectra changes in Sco X-1 \citep{Titarchuk+etal+2014}.
Since the energy spectra of the bottom NB and the bottom FB are not significantly different, the timing signatures can be used as an additional identification of the $Z$ branch.

\subsection{The identification of kHz QPOs} \label{subsec:d3}

The $\sim$ 40 Hz HBOs and $\sim$ 800 Hz kHz QPOs are observed simultaneously on the HB of Sco X-1 up to $\sim$ 60 keV. The centroid frequencies of both the HBOs and kHz QPOs increase from the top to the bottom of the HB, which are consistent with those reported by \cite{Klis+etal+1997} and \cite{Bu+etal+2015}, featuring a moving inward disk-corona geometry. Similar to NBOs, the HBO and kHz QPO frequencies do not change significantly with photon energy, while their rms increase with photon energy, suggesting a non-thermal origin for both the HBOs and kHz QPOs (see also \cite{Klis+etal+1996}, \cite{Lin+etal+2011}, \cite{Motta+etal+2017}).

Despite the fact that LFQPOs have been studied for several decades, their origin is still uncovered. The relativistic Lense Thirring precession seems to be, to date, the most promising mechanism for the existence of certain LFQPOs both in NS and BH LMXBs. In this model, an HBO with $\sim$ 40 Hz up to 60 keV can be explained assuming that the QPO is caused by the precession of the hot inner flow. However, no solid conclusion can be given until the time lag of HBO is studied, which unfortunately can not be done at the moment from our data due to the limited photon counts. Longer and more frequency observations are suggested for further work.

Twin kHz QPO peaks with an upper peak of $\sim$ 820 -- 1150 Hz and a lower peak $\sim$ 540 -- 860 Hz have been found in Sco X-1 with the $RXTE$ data (\citealt{Klis+etal+1996}, \citealt{Klis+etal+1997}, \citealt{Yu+etal+2001}, \citealt{Mendez+Klis+2000}, \citealt{Belloni+etal+2005}), and towards the edges of the observed frequency range, the kHz QPO also occurs alone (\citealt{Klis+2006}). In our work, we detected a twin kHz QPOs with a lower peak of $\sim$ 550 Hz (3.2$\sigma$) and a higher peak of $\sim$ 824 Hz (3.0$\sigma$) and the single kHz QPOs $\sim$ 800 Hz with a significance larger than 3.0$\sigma$ on the HB, shown as in Table~\ref{Tab:tablekHzQPO} and Fig.~\ref{Fig:kHzQPO}. If we consider the correlation of the frequencies between the HBOs and the upper kHz QPOs (\citealt{Klis+etal+1997}), an HBO with frequency of $\sim$ 35 -- 40 Hz corresponds to an upper kHz QPOs of $\sim$ 800 -- 900 Hz. Therefore, the $\sim$ 800 Hz single kHz QPOs detected by $Insight$-HXMT should be the upper peak of the twin kHz QPOs.
This is the first decisive detection of kHz QPOs in 20 -- 60 keV. As it is difficult for thermal processes to produce photons with such high energies, the kHz QPOs we detected correspond quite probably to the dynamics in the non-thermal emission regions. If the frequencies represent the Keplerian frequencies of the inner most accretion disk (\citealt{Miller+etal+1998}), our results imply that the inner most region of the accretion disk is non-thermal in nature. The detailed  discussions about the properties and the origin of the kHz QPOs in out of the scope of this paper. We leave them to a future paper.

\section{SUMMARY} \label{sec:conclusion}

In this work, we presented a broad energy band timing analysis of Sco X-1 observed with \emph{Insight}-HXMT from July 2017 to August 2018. The energy-dependence of QPO properties evolving along the position of the $Z$-track are investigated in detail. The main results are summarized as what follows:

1) A complete $Z$-track HID is produced, showing a $v$-like shape.

2) $\sim$ 6 Hz NBOs in the lower part of the normal branch and $\sim$ 16 Hz FBOs in the beginning part of the flaring branch are obtained by LE and ME. $\sim$ 40 Hz HBOs with $\sim$ 800 Hz kHz QPOs are detected simultaneously by ME and HE. For all types of QPOs, the centroid frequencies do not vary with energy within 1 $\sigma$ uncertainties, while the fractional rms increase with energy, featuring a non-thermal origin.

3) The centroid frequencies of both the HBOs and kHz QPOs increase along the $Z$-track from the top to the bottom of the HB, suggesting a moving inward disk-corona geometry.

4) NBO phase lags show soft lags increasing with energy, suggesting a non-homogeneous corona geometry.

5) We detected the $\sim$ 800 Hz KHz QPOs in 20 -- 60 keV, which is the first decisive detection of kHz QPO above 20 keV and implies that the inner most region of the accretion disk is non-thermal in nature.

\acknowledgements {

This work made use of the data from the $Insight$-HXMT mission, a project funded by China National Space Administration (CNSA) and the Chinese Academy of Sciences (CAS). The authors thank supports from the National Program on Key Research and Development Project (Grant No. 2016YFA0400801), the Bureau of International Cooperation, Chinese Academy of Sciences (GJHZ1864), the Strategic Pioneer Program on Space Science, Chinese Academy of Sciences (Grant No. XDA15310300) and the National Natural Science Foundation of China (Grant No. 11733009, 11673023, U1838111 and U1838115).

}

\bibliography{biblio}

\end{document}